\documentclass[iop]{emulateapj}
\usepackage{graphicx}

\usepackage{color}

\usepackage{epstopdf}
\usepackage{amsmath}

\def\ltsima{$\; \buildrel < \over \sim \;$}
\def\simlt{\lower.5ex\hbox{\ltsima}}
\def\gtsima{$\; \buildrel > \over \sim \;$}
\def\simgt{\lower.5ex\hbox{\gtsima}}
\def\obs{{\rm obs}}
\def\inf{{\rm ap}}

\newcommand\lsim{\mathrel{\rlap{\lower4pt\hbox{\hskip1pt$\sim$}}
\raise1pt\hbox{$<$}}}
\newcommand\gsim{\mathrel{\rlap{\lower4pt\hbox{\hskip1pt$\sim$}}
\raise1pt\hbox{$>$}}}

\usepackage{color}

\shorttitle{Confusing binaries in the Galactic Center}
\shortauthors{S. Naoz et al.}

\begin{document}

\title{Confusing binaries: the role of stellar binaries in biasing  disk properties  in the Galactic Center}

\author{Smadar Naoz\altaffilmark{1,2},  Andrea M.~ Ghez\altaffilmark{1},  Aurelien Hees\altaffilmark{1}, Tuan Do\altaffilmark{1},  Gunther Witzel\altaffilmark{1}, Jessica R.~Lu\altaffilmark{3} \\}
\affil{$^1$Department of Physics and Astronomy, University of California, Los Angeles, CA 90095, USA}
\affil{$^2$Mani L. Bhaumik Institute for Theoretical Physics, Department of Physics and Astronomy, UCLA, Los Angeles,
CA 90095}
\affil{$^3$Astronomy Department, University of California, Berkeley, CA 94720, USA}


\email{snaoz@astro.ucla.edu}

\begin{abstract}
The population of young stars near the Supermassive black hole (SMBH) in the Galactic Center (GC) has presented an unexpected challenge to theories of star formation. Kinematics measurements of these stars have revealed a stellar disk structure (with an apparent $20\%$ disk membership) that has provided important clues to the origin of these mysterious young stars.   However many of the apparent disk properties are difficult to explain, including the low disk membership fraction and the high eccentricities, given the youth of this population. Thus far, all efforts to derive the properties of this disk have made the simplifying assumption that stars at the GC are single stars.
Nevertheless, stellar binaries are prevalent in our Galaxy, and recent investigations suggested that they may also be abundant in the Galactic Center. Here we show that binaries in the disk can largely alter the apparent orbital properties of the disk. The motion of binary members around each other adds a velocity component, which can be comparable to the magnitude of the velocity around the SMBH in the GC. Thus, neglecting the contribution of binaries can significantly vary the inferred stars' orbital properties. While the disk orientation is unaffected the apparent disk's 2D width will be increased to about $11.2^\circ$, similar to the observed width.  
For a population of stars orbiting the SMBH with zero eccentricity,  unaccounted for binaries will create a wide apparent eccentricity distribution with an average of $0.23$.   
This is consistent with the observed average eccentricity of the stars' in the disk.   We suggest that this high eccentricity value, which poses a theoretical challenge, may be an artifact of binary stars.
Finally, our results suggest that the actual disk membership might be significantly higher than the one inferred by observations that ignore the contribution of binaries, alleviating another theoretical challenge. 
\end{abstract}

\maketitle

\section{introduction}

Recent observations in the Galactic Center (GC) revealed a population of young stars, which is consistent with a star formation episode $\sim  4-6$~Myr ago \citep[the young nuclear cluster, YNC, e.g.,][]{Lu+09,Bartko+10,Do+13kin,Do+130.5pc,Feldmeier+15}.
These young stars appear to have two distinct kinematic structures around the Supermassive Black Hole (SMBH). Closer to the SMBH ($\lsim 0.04$~pc) the stars have an isotropic distribution with high eccentricities \citep[following a thermal distribution][]{Gillessen+17} and may be slightly older as the highest mass stars are B stars \citep{Ghez+03}. These stars are the so called S-star cluster 
 \citep[e.g.,][]{Schodel+03,Ghez+05,Eisenhauer+05,Ghez+08,Gillessen+09,Bartko+10,Yelda+14}. 
 Between $\sim 0.04$~pc and $0.5$~pc, roughly $3/4$ of the stars are old, late-type giants and the remaining $1/4$ are young stars, including many massive Wolf-Rayet stars  \citep{Levin+03,Genzel+03,Eis+05,Paumard+06,Lu+09,Bartko+09,Yelda+14}.

Over time, the kinematic picture is becoming more precise as it has become possible to measure accelerations on the plane of the sky.   Recently, \citet{Yelda+14}, using high precision kinematic measurements and modeling of $116$ young stars in the inner $\sim 1$~pc of the GC,  suggested that the stellar disk is composed of  $\sim 20\%$ of the stars sampled between $\sim 0.04$~pc and $0.5$~pc.
  The off-disk stellar population is less constrained and seems to be extended beyond $\sim 0.1$~pc ($\sim 3.2''$). The existence of additional substructure in this regime has been suggested (for example a warped disk), but it seems to have low significance \citep{Bartko+09}.
Dynamical modeling of the stars in the disk yields an average stellar eccentricity of $\sim 0.3$ \citep{Yelda+14}. Moreover, the stars in the disk seem to follow a moderately top-heavy mass function \citep{Bartko+10,Lu+13}.  In addition, the B stars in the disk have on average, similar kinematic properties as the more massive O and WR stars, suggesting a common star formation event \citep{Lu+13}. 
 
The presence of the disk and its proprieties provide an important constraint to the origin of the young stars, which are hard to form in the inhospitable environment around the SMBH \citep[e.g.,][]{Levin07}.  For example, it was suggested that the young S-star cluster might have formed further out in the disk and migrated in \citep[e.g.,][]{Levin+03,Nayakshin06,Levin07,Alexander+08}. Note, that there are many other ideas in the literature of the origin of the young stars in the  GC, varying from breaking up binaries \citep[e.g.,][]{Hills88,Yu+03,Perets09}, to collisions and mergers \citep[e.g.,][]{Ginsburg07,AP12,Stephan+16}, and others \citep[e.g.,][]{Perets07,Hobbs+09,Perets+09}.   However, the kinematic substructure of the YNC, mentioned above, suggests that the cluster, and thus the young stars, formed in-situ rather than far from the SMBH 
\citep[e.g.,][]{Berukoff+06,Lu+09,Yelda+14,Stostad+15,Feldmeier+15}.

The stellar membership estimations and the modeling of the disk's properties are based on multiple years of astrometric measurements, but typically only use {\it one} radial velocity ($V_z$) measurement  \citep[e.g.,][]{Bartko+09,Yelda+14}.
However, if at least some of the stars in the disk are in a binary configuration, the associate z-component velocity may be misleading, resulting in poor disk membership interpretation, and with larger apparent eccentricity around the SMBH than the physical one. Interestingly, among the stars with detected accelerations on the plane of the sky, which generally have the most robust disk membership estimate, the photometrically identified binary \citep{Martins+06} IRS 16SW has the lowest disk membership probability \citep{Yelda+14}.

Recent observations have suggested that binaries are prevalent in our Galaxy \citep[$\gsim 70\%$ for ABO spectral type stars, e.g.,][]{Raghavan+10}. Thus, on face value, the binary fraction should be large among the young stars in the GC as well. So far, there have been three confirmed binaries in the inner $\sim0.2$~pc of the GC. The first confirmed binary (IRS 16SW) is an equal-mass binary ($50$~M$_\odot$) at a projected distance estimated as $\sim0.05$~pc with a period of 19.5 days \citep{Ott+99,Martins+06,Rafelski+07}. Recently, \citet{Pfuhl+13} discovered two additional binaries, an eclipsing Wolf-Rayet binary with a period of 2.3 days, and a long-period binary with an eccentricity of 0.3 and a period of 224 days. Both of these binaries are estimated to be at only $\sim0.1$~pc from the SMBH.  These observational studies suggest that the total massive binary fraction in the Young nuclear star cluster is comparable to the galactic one  \citep[e.g.,][]{Ott+99,Rafelski+07}.  Recently \citet{Stephan+16} showed that the binary fraction in the nuclear star cluster might be as high as $70\%$, compared to the initial binary fraction, following a star formation episode that took place in that region a few million years ago \citep[e.g.,][]{Lu+13}. 
Furthermore, other studies focusing on the overabundance of observed X-ray sources in the central 1pc suggests that compact binaries, involving stellar-mass BHs and neutron stars, may reside there in larger than average numbers \citep{Muno+05}. Together, these studies suggest that binaries are likely to be prevalent at the GC.

 \begin{figure*}
\centering
\begin{minipage}[t][][b]{\textwidth}
\centering
\hspace{-4.5cm}
\begin{minipage}[b]{.32\textwidth}
\centering
\includegraphics[scale=.49]{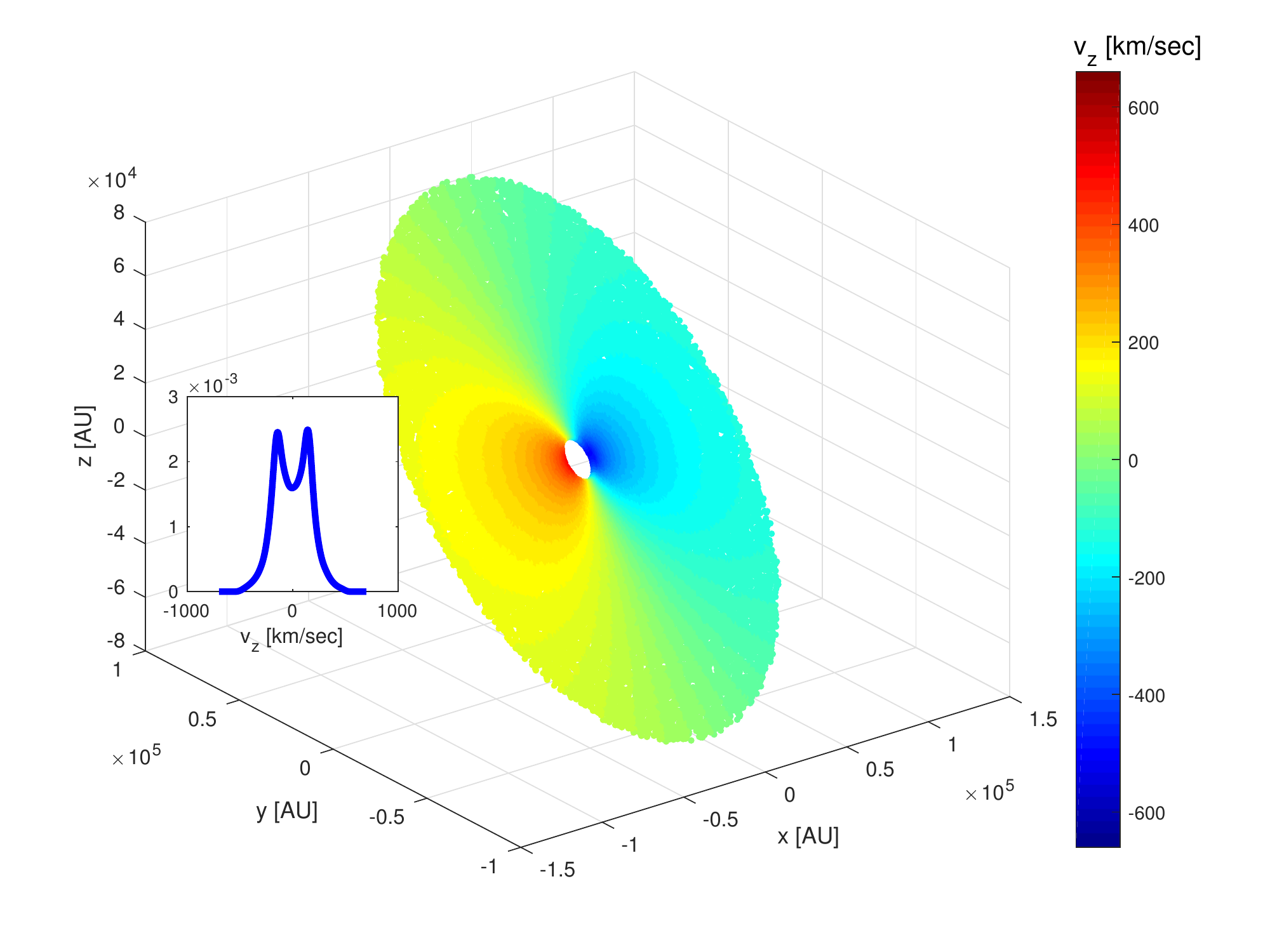}
\end{minipage}\hspace{3.8cm}
\begin{minipage}[b]{.32\textwidth}
\centering
\includegraphics[scale=.49]{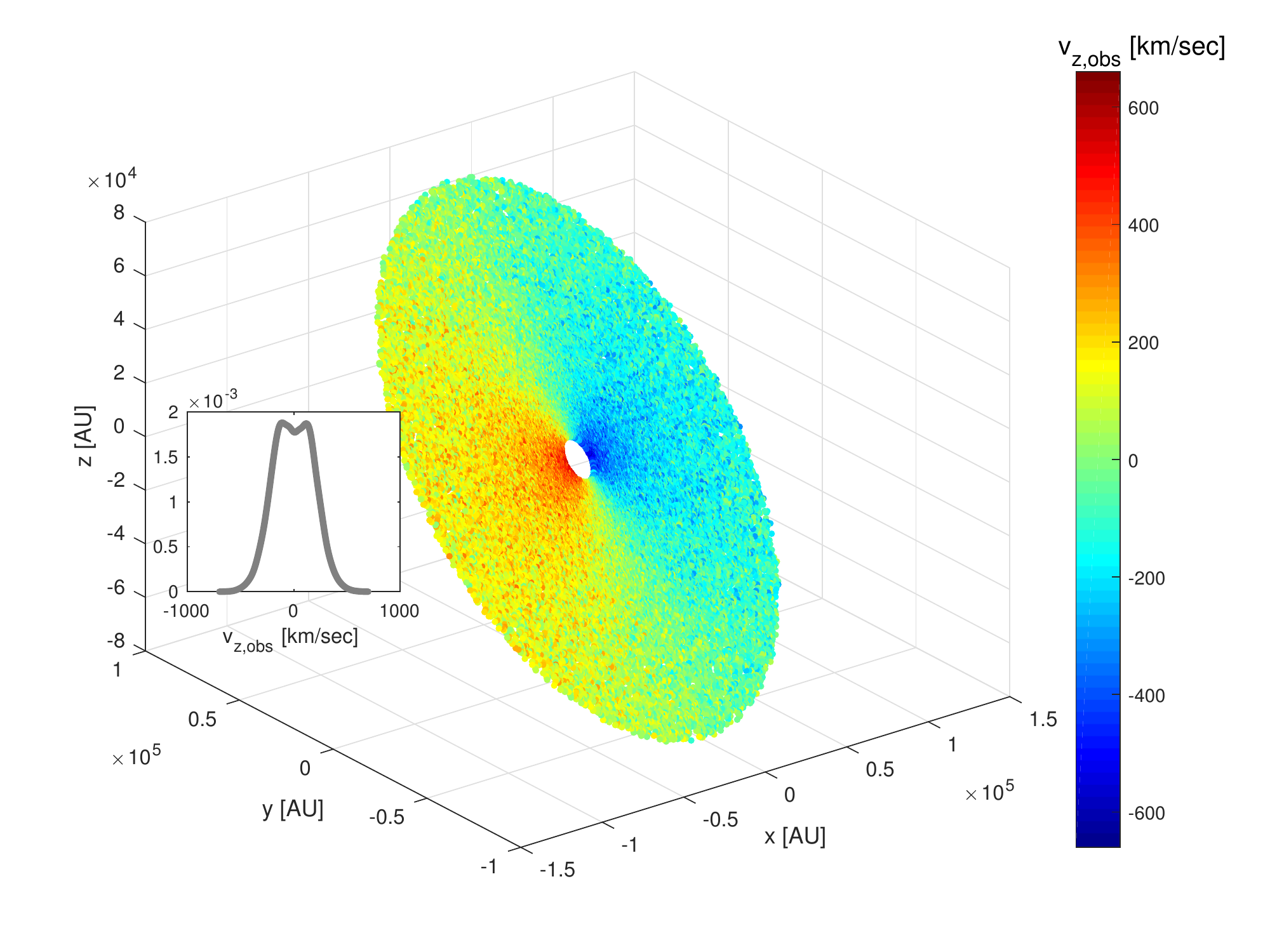}
\end{minipage}
\end{minipage}\par\medskip 
\begin{minipage}[b]{\textwidth}
\centering
\includegraphics[scale=.52]{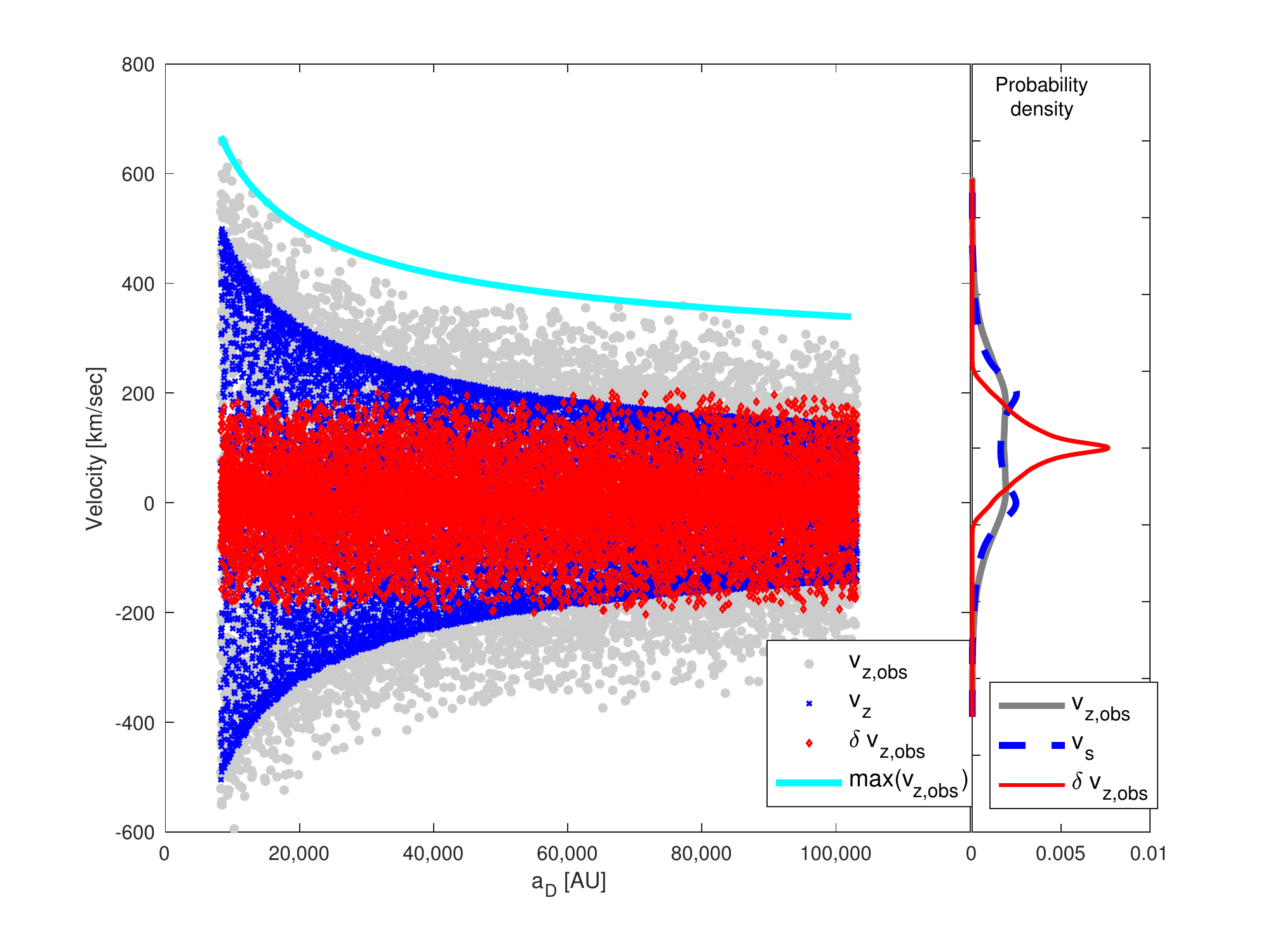}
\end{minipage}
\qquad\caption{ {\it Top panels}: 3D representation of the disk. {\it Top left panel}: stars in the disk, without accounting for the binaries' affect on the velocity. The inset shows the distribution of the the $v_z$ component for the binaries' center of mass motion around the SMBH. 
 {\it Top right panel}: stars in the disk while accounting for binaries affect on the velocities. 
 The color code shows the observed z-component velocities for binary stars, i.e., $v_{z,{\rm obs}}=v_z+\delta v_{z,{\rm obs}}$. The inset shows the distribution of the z-component velocity of these binary stars around the SMBH.  {\it Bottom panel} shows the $z$-component velocity as a function of the stellar disk's semi-major axis $a_D$. We show the input velocity $v_z$ (blue), the induced wobble $\delta v_{z, {\rm obs}}$ (red) and the observed part $v_{z,\rm obs}=v_z +\delta v_{z, {\rm obs}}$. For illustrative purposes we show only $20\%$  from the Monte-Carlo simulations. The cyan line shows the maximum value of $v_{z,\rm obs}$. We show the associated probability densities of these quantities in the right bottom panel.   Note that unless said explicitly, we consider the $e_D=0$ case.  }
\label{fig:Diskvel}
\hspace{1cm}
\vspace{0.5cm}
\end{figure*}

 Several groups have begun to explore the dynamical effects of binaries (both stellar and compact objects) in shaping the physical properties of stellar distribution in the GC. For example, binaries are invoked to explain some long-standing observational puzzles, such as hypervelocity stars, the young stars in the S-cluster, the dark cusp, etc., \citep[e.g., ][]{Hills88,Yu+03,Antonini+10,OLeary+09,Perets+09,AH09,AP12,AP13,Phifer+13,Prodan+15,Witzel+14,Witzel+17,Stephan+16}. Furthermore, it has been suggested that compact object binaries in the GC are a potential source of gravitational wave emission \citep[e.g.,][]{OLeary+09,AP12,Prodan+15,Hoang+17}.

Within the vicinity of a SMBH, the members of a stable binary have a tighter orbital configuration than the orbit of their mutual center of mass around the SMBH. In such a system, gravitational perturbations from the SMBH can induce large eccentricities on the binary orbit, which can cause the binary members to merge \citep[see for review of the dynamics][]{Naoz16}. This coalescence may form a new star that can look like the G2 and G1 objects \citep{Phifer+13,Prodan+15,Witzel+14,Witzel+17,Stephan+16}, which may eventually become blue stragglers \citep[e.g.,][]{NF}.

Here we suggest that some of the puzzling observations associated with the stellar disk may arise from neglecting the contribution of binaries to the kinematic measurements. {Ignoring the contribution of binaries is known to cause an overestimation of the dynamical mass of star clusters \citep[e.g.][]{Kouwenhoven+08}.}
We demonstrate the importance of including binaries in the GC kinematic modeling by adopting a thin stellar disk with a population of binaries. We focus on the effects in inferring the disk's properties as a result of neglecting the presence of these binaries. In particular, we show that ignoring the motion of binaries may lead to reduced disk membership and an increased apparent eccentricity and semi-major axis of the stars in the disk. 
We present our initial conditions in Section \ref{sec:ICs} and present the spirant biases and disk properties in Section \ref{sec:ap}, and finally, discuss our results and implication in Section \ref{sec:dis}.

\section{Initial conditions for disk  Binaries  }\label{sec:ICs} 

 \begin{figure}
\hspace{-2.8cm}
\centering
\begin{minipage}[t][][b]{.4\textwidth}
\centering
\includegraphics[scale=.32]{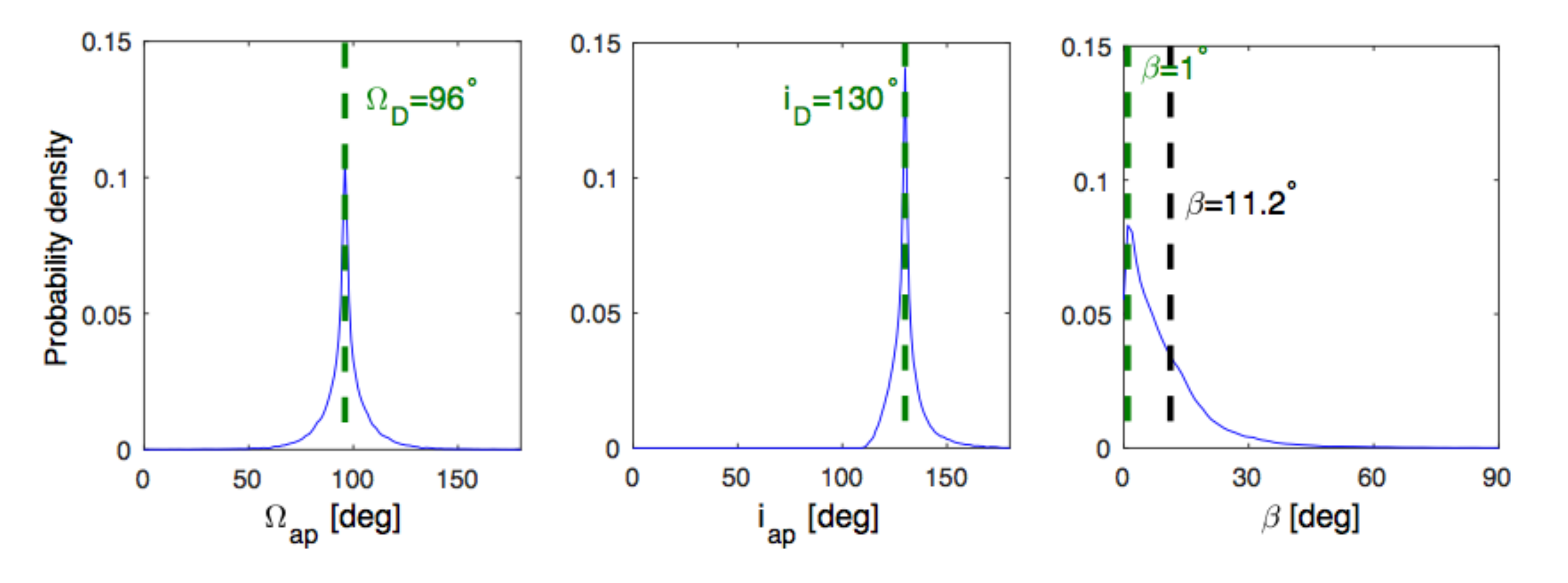}
\end{minipage}\par\medskip 
\begin{minipage}[b]{.4\textwidth}
\hspace{-1.4cm}
\includegraphics[scale=.38]{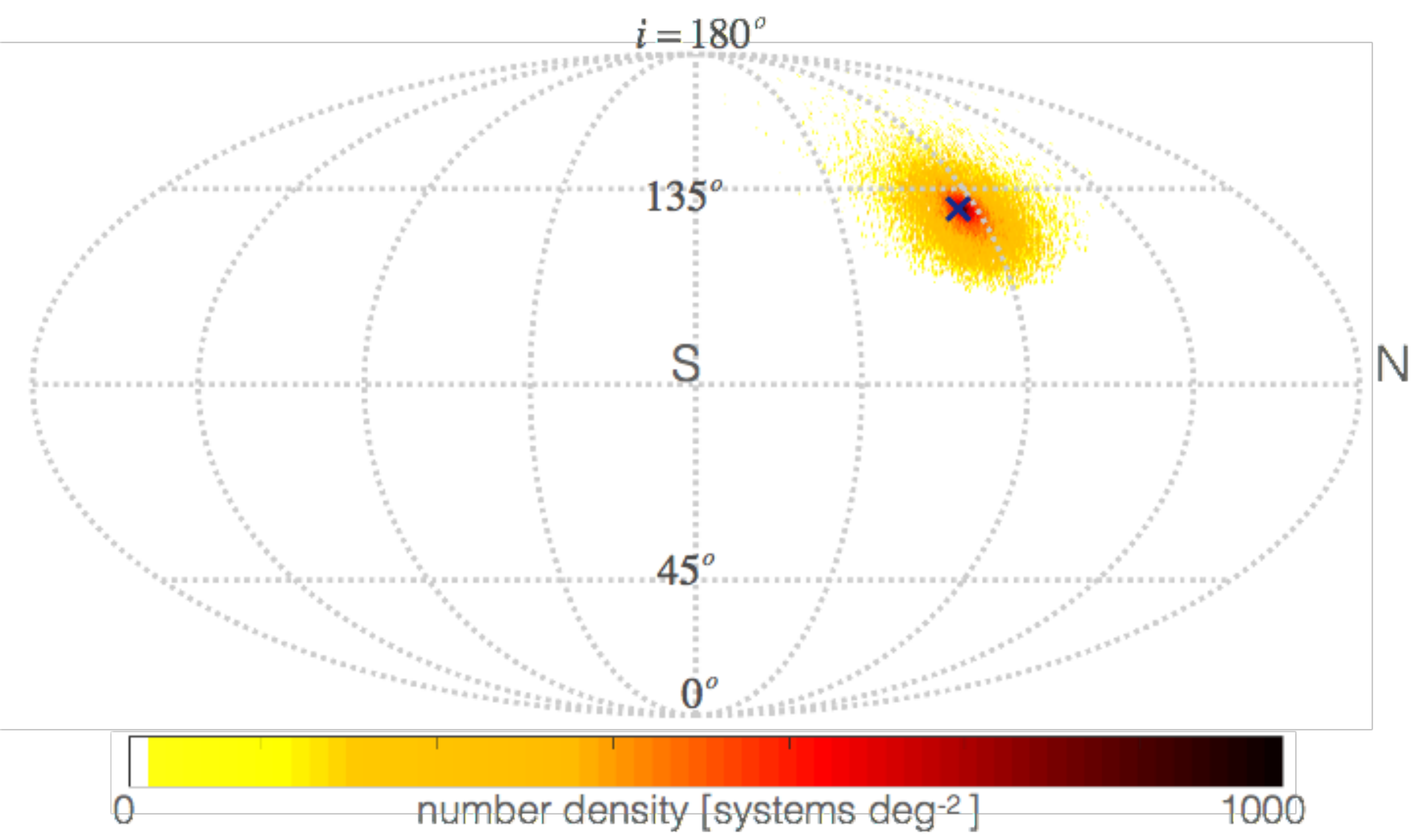}
\end{minipage}
\qquad\caption{ {\it Top row}: The density probability of the apparent orbital parameters as a result of the binary wobble. We show the (right from left) the apparent  ascending nodes ($\Omega_\inf$), inclination, ($i_\inf$),  and the quantity $\beta$. No change in the parameters yields a zero tilt angle. {The green dashed lines in each panel mark the peak of the distribution. The black dashed line represents the $\beta$ angle associated with $68\%$ of the population (see text). } {\it Bottom panel}: Apparent  sky projection of the disk in an equal area Mollweide projection.
The direction of the normal vector is described by the inclination ($i_\inf$), depicts here by horizontal lines spaced $45^\circ$ apart) and the angle to the ascending node ($\Omega_\inf$) longitudinal lines spaced  $45^\circ$ apart, with the line marked N representing  $0^\circ$). The razor thin initial disk was set to be at $(i_D,\Omega_D)=(130^\circ,96^\circ)$.
 The color shows the density of systems in $0.2$ square degree pixel area. The light blue X marks the maximum density peak disk value, which corresponds to $i_D$ and $\Omega_D$.  We also note that  some systems do exist in the left half sphere of the sky-projection plot, but the density is very low corresponding to light yellow in the color palette.}
\label{fig:angles}
\hspace{1cm}
\end{figure}

We present a proof-of-concept calculation of the effect of the binary motion on the apparent orbital parameters. We assume a razor thin stellar disk around a $4\times 10^6$~M$_\odot$ SMBH \citep[e.g.,][]{Ghez+08,Boehle+16,Gillessen+17}. We model the disk extending from $0.04$~pc to $0.5$~pc around the SMBH. The semi-major axis of the orbit around the SMBH ($a_2$) is chosen to have a Bahcall-Wolf-like distribution with the number density $n(r)\sim r^{-2}$ \citep[as supported by observations, e.g.,][]{Lu+09,Stostad+15}. Note that the number of stars in a shell can be written as $dN=4\pi r^2 n(r)dr$, and thus the choice of the semi-major axis around the SMBH follows a uniform distribution in $a_2$ (i.e., constant $dN/dr$) between $0.04$~pc and $0.5$~pc. We adopt \citet{Yelda+14} orbital parameters for the disk location on the Sky. Specifically, we set $i_D=130^\circ$ and $\Omega_D=96^\circ$. The angle $i_D$ is the inclination with respect to the observer, such that $\cos i_D=h_z/h_D$, where $h_z$ is the z-component of the angular momentum of a star ${\bf h}_D$ and $h_D$ is the magnitude of the stellar orbital angular momentum around the SMBH. {The $z$ axis is along the line of sight} and the observer is located at $z=-8.32$~kpc \citep[e.g.,][]{Gillessen+17}.  
In the first simulation, each binary star system is assigned to be on a circular velocity around the SMBH (i.e., $e_D=0$ for all the orbits).  In a second simulations, the binaries are put on orbits around the SMBH with $e_D=0.3$. The nominal, default, case we consider throughout the paper is the case for $e_D=0$, and we will refer to the eccentric case for comparison in our results.  We model a total of $80,000$ systems for two disk's 
eccentricity choices ($40,000$ per choice of disk's eccentricity) and for each system use a uniform distribution to chose (1) the argument of periapsis of the orbit around the SMBH, $\omega$ ($0^\circ$ to $360^\circ$) and (2) the mean anomaly of the orbit around the SMBH ($0^\circ$ to $180^\circ$), from which we find the true anomaly for the circular and eccentric cases. 
 Given these orbital parameter we can find the location (${\bf r}_D=(x,y,z)$) and velocity (${\bf v}_D=(v_{x},v_{y},v_{z})$) of the stars in the disk \citep[e.g.,][]{MD00}.  We note that compared to the \citet{MD00} transformation, the observation convention is such that $x\to y$ and $y\to -x$. This convention corresponds to having the origin of the longtime of ascending nodes ($\Omega$) in the north, corresponding to positive $x$ \citep[e.g.,][]{Ghez+05}.  The top left panel in Figure \ref{fig:Diskvel} shows the 3D representation of the disk, color coded by the z-component of the velocities around the SMBH, $v_z$.

We assume that every star in the disk is actually a $m=10$~M$_\odot$ star with a binary companion, with a mass ratio $q$.
 The mass ratio was chosen from a uniform distribution between $0$ and $1$ {to match the distributions of the Galactic O star population \citep[e.g.,][]{Sana+11,Sana+12}}. We assume that all the binaries are on a circular orbit with a semi-major axis of $0.1$~AU \citep[e.g.,][Ginsburg et al in prep.]{Li+17,Chu+17}, which sets the binary period to be $P=3.6/\sqrt{1+q}$~days. {Larger separations are more sensitive to eccentricity excitations due to the Eccentric Kozai-Lidov (EKL) mechanism as they yield shorter EKL timescale compared to General Relativity\footnote{{Note that comparable precession timescale between General Relativity and EKL can still cause eccentricity excitations in the form of resonant behavior, \citep[e.g.,][]{Naoz+13} thus we have chosen a much larger difference in the timescales.} }, which can lead to merging the two binary members \citep[e.g.,][]{Stephan+16,Li+17}. Furthermore, gravitational perturbations from fly-by stars that eventually can unbind the binary \citep[e.g.,][]{Binney+87}. Thus, we adopt a hard binary \citep[e.g.,][]{Quinlan96} with $0.1$~AU separation as a proof-of-concept. }  {The induced observed wobble of the primary at the z-direction}\footnote{{Note that velocity wobble due to the stellar companion may also be observable because of the flat mass ratio distribution. For simplicity, we ignore this contribution, and assume that these companions are fainter than the primaries.}}, due to the binary, (either positive or negative) is then  (for a binary on a circular orbit around it's center of mass)
\begin{equation}\label{eq:dv}
\delta v_{z,\rm obs}^3 =  2\pi G \frac{q^3}{(1+q)^2} \frac{m}{P}\sin^3 i_{\rm bin} \cos^3 f_{\rm bin} \ ,
\end{equation}
where $i_{\rm bin}$ is the binary inclination, drawn from a isotropic distribution between $0-180^\circ$ {(i.e,. uniform in $\cos i_{\rm bin}$)}, $f_{\rm bin}$ is the true anomy of the binary, and  $G$ is the gravitational constant. 
Each orbit on the disk with a velocity vector ${\bf v}_D$ will have an additional observed velocity component along the z direction $\delta {\bf v}_z$. In other words the observed velocity will be ${\bf v}_{\rm obs}={\bf v}_D+\delta {\bf v}_{z,\rm obs}$.  

 Note that our choose of a random binary mass ratio, binary phase and inclination, reduces the impact of the binaries. On the other hand we fixed the binary separation and assumed a circular binary. Allowing for a distribution of binary separations might increase in some cases and decrease in others the wobble value.Thus, we do not expect a qualitatively change the results of this proof-of-concept calculations. 

The wobble velocity $\delta {\bf v}_{z,\rm obs}$ will be at the order of ${\bf v}_D$ at characteristic distances from the SMBH
 \begin{equation}
a_{D,c}\sim a_1\frac{M}{m} \frac{1+q}{q^2}\frac{1}{\sin i_{\rm bin}} \sim 8\times 10^4 \left(\frac{M}{4\times 10^6 {\rm M}_\odot}\right)~{\rm AU}
\end{equation}
 For example a typical value for a star in the disk is $\sim 200$~km~sec$^{-1}$ (see inset in the left panel of Figure  \ref{fig:Diskvel}), and a typical wobble for these massive binaries are $\sim 100$~km~sec$^{-1}$ {see Figure \ref{fig:Diskvel}}. Therefore, the resulted observed z-component velocity spread on somewhat larger range and has different velocity distribution (see Figure \ref{fig:Diskvel}, top right panel). Binaries that are closer to the SMBH will be less sensitive to the additional velocity induced by the binary. As depicted in Figure  \ref{fig:Diskvel}, further away from the SMBH the Keplerian velocity around the SMBH may be comparable to that of the binary wobble and thus can greatly affect the inferred orbital properties.

\section{Disk property biases induced by Binaries }\label{sec:ap}

Given this ``new" observed velocity component we can find the ``observed" (apparent)  orbital parameters. In other words, treating a binary system as a single star can result in inferring  orbital parameters that are inconsistent with the physical ones.  Specifically, we are interested in  $\Omega_{\inf}$ and $i_{\inf}$, where the subscript ``ap" stand for apparent. We calculate the apparent angular momentum ${\bf h}_\inf$ and thus, the angles can be easily found using the following relations   \citep[e.g.,][]{MD00}:
\begin{equation}
\cos i_\inf = \frac{h_{z,\inf}}{h_\inf}  \ , 
\end{equation}
and 
\begin{equation}
\cos \Omega_\inf =  \frac{-h_{y,\inf}}{h_\inf\sin i_\inf} \quad {\rm and} \quad \sin \Omega_\inf = \frac{h_{x,\inf}}{h_\inf\sin i_\inf} \ .
\end{equation}
We show the probability density of these angles in Figure \ref{fig:angles} top panels\footnote{The probability density is calculated based on a normal kernel function, and is evaluated at equally-spaced points of the relevant angle.}. As expected the resulted angles have a wide distribution that peaks at the disk value, i.e., $i_D=130^\circ$ and $\Omega_D=96^\circ$. These values are close to the median of the distributions. However, due to the long tail the averages are different. Specifically, we find, $<i_\inf >\sim 130.6^\circ$ and $<\Omega_\inf>\sim95.6^\circ$ with standard deviations of $\sim 7.8^\circ$ and $\sim 15^\circ$ respectively.   These angles define the location of the disk on a sky, and in Figure \ref{fig:angles}, bottom panel,  we show the apparent sky projection, color coded with the density of stars from our simulations. We define the density as the number of stars in a degree area of $i_{\inf}\times\Omega_\inf$. Note that the maximum density takes place at $i_D$ and $\Omega_D$, marked by light blue $X$ in the bottom Figure of  \ref{fig:angles}, as indicated in the probability distribution in the top of Figure  \ref{fig:angles}. {Note the slight ``shadow"  near the $X$ is a result of over-density of slightly dark pixels and it corresponds to roughly $1\sigma$ of disk membership (see below).} 
 The spread around the razor thin disk is the result of the relative motion of the binaries around each-other. Interestingly, the density of the apparent disk members  around the maximum, projected on the sky, is not dissimilar from the observations \citep[see for example][figure 10]{Yelda+14}.

 \begin{figure*}
 \centering
 \includegraphics[width=\linewidth ]{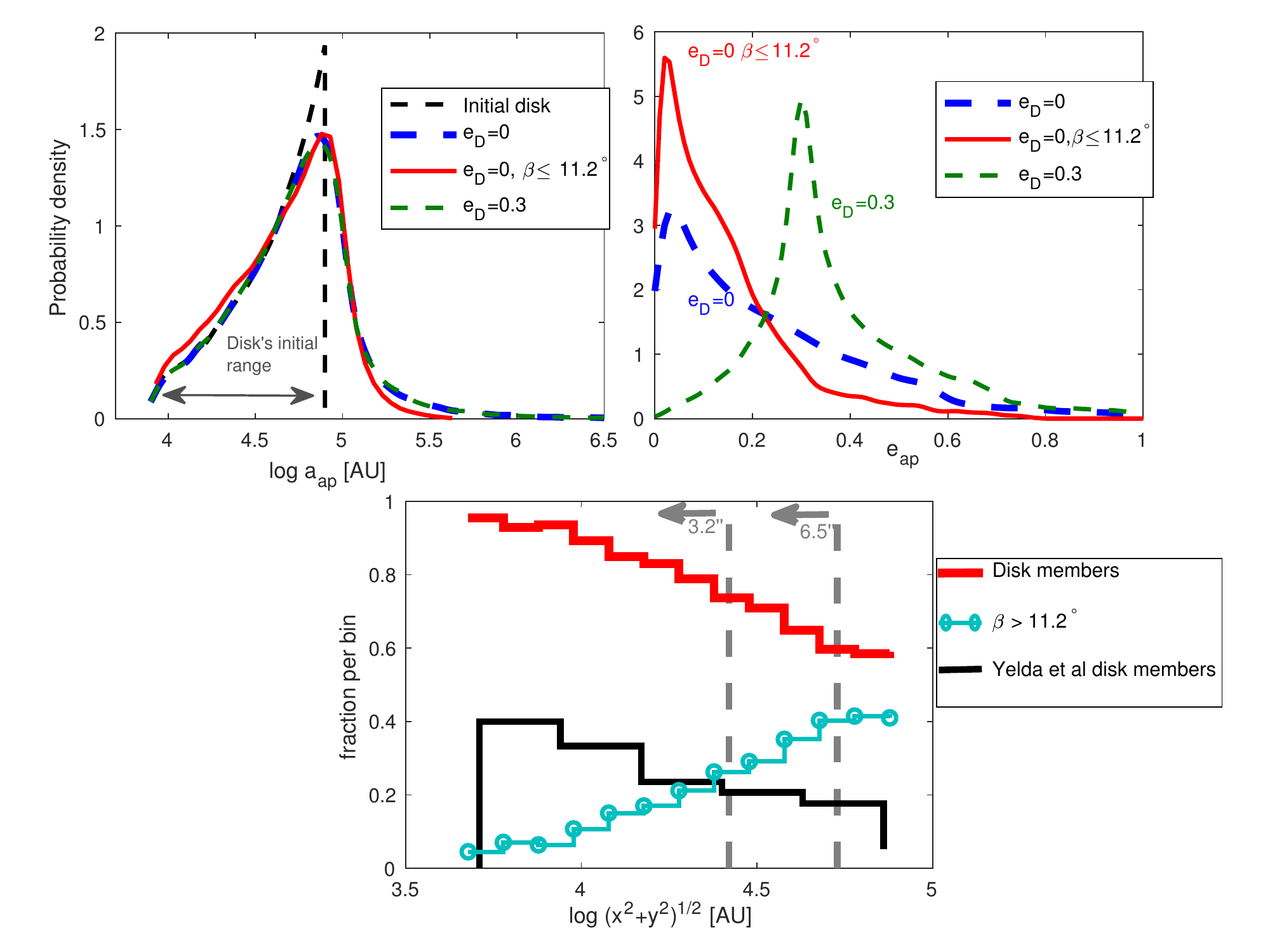}
\caption{{\it Top raw}: {The density probability of the apparent semi-major axis ($a_\inf$), left panel, and the apparent eccentricity ($e_\inf$), right panel. We show the two cases associated with two different initial eccentricities for each star in the disk, $e_D=0$ (blue dashed lines) and $e_D=0.3$ (green dashed lines). The two initial eccentricity cases give the same density probabilities for the angles in Figure \ref{fig:angles}. The dashed gray line in the left panel shows the probability density of the initial disk and it's range. The solid red lines show the  disk members' eccentricity (right) and $a_\inf$ (left) density probability for the $e_D=0$ run. Disk members are defined as those with $\beta\leq11.2^\circ$ (i.e., $\beta_{68}$). {\it Bottom panel}: The fraction of stars in the disk (red line) represented as stars with $\beta\leq 11.2^\circ$ (solid line). Stars with $\beta>11.2^\circ$ are inferred to lay outside the disk (solid, cyan line). 
The black line shows the disk members estimated from the 116 stars from \citet{Yelda+14}. The observed disk membership is normalized such that  $20\%$ of the stars reside in the disk (see text). } Overplotted are the sky projected radius bins at $3.2''$ and $6.5''$ when assuming a distance to the SMBH of $8.32$~kpc. {Note that the observed disk membership  per projected distance bin and our calculations show the same trend, with a $\sim 2.5$ difference. This similarity implies that the disk membership might be as high as $\sim 50\%$ instead of the inferred $\sim 20\%$ fraction achieved while ignoring binaries. }  } 
  \label{fig:SMAEcc} 
\end{figure*}

We define a quantity
 \begin{equation}\label{eq:b} 
\beta=\sqrt{(\Omega_\inf -\Omega_D)^2+(i_\inf-i_D)^2} \ , 
\end{equation}
 which can be used as a proxy for disk membership (see below).  In Figure \ref{fig:angles}, top right panel we show probability density of $\beta$ which peaks at $1^\circ$. The deviation from a peak at $\beta=0^\circ$ can be understood from the asymmetry of the inclination distribution.  The
 In terms of the number systems, the peak of this distribution for $i_\inf$ is about a factor of two smaller than that for $\Omega_\inf$. Combing $i_\inf$ and $\Omega_\inf$ yields $\beta$. 
 We found that the average $<\beta>\sim 10.1^\circ$ with a standard divination of $13.5^\circ$. 
{
Because the peak distributions in $\Omega_\inf$ and $i_\inf$ are similar to the initial disk input, the apparent peak density of stars in the sky projected disk (bottom panel in Figure \ref{fig:angles}) is at the initial disk input (marked by the light blue X). 
  Below we adopt the disk membership by estimating the critical angle $\beta_{68}$ at which $68\%$ of the systems lay within it around the disk,  which corresponds to $\beta_{68}=11.2^\circ$. The $68\%$ from the initial disk's input (i.e., $\beta=0$) is shown in the vertical dashed black line in Figure \ref{fig:angles}. }
 
We also calculate the apparent eccentricity in the disk using Kepler relation, i.e.,
\begin{equation}\label{eq:e}
e_\inf=\sqrt{1-\frac{h_\inf^2}{G (M +m(1+q)) a_\inf}} \ ,
\end{equation}
where $M$ is the SMBH mass and $a_\inf$ is found from the following Kepler relation 
\begin{equation}\label{eq:a}
\frac{1}{a_\inf}=\frac{2}{r_D} - \frac{v_\obs^2}{  G (M +m(1+q))}  \ .
\end{equation}
We show the distribution of the apparent semi-major axis and eccentricity in Figure \ref{fig:SMAEcc}, top left and right panels, respectively. As depicted in this Figure, closer to the SMBH, the apparent semi-major axis roughly follows the razor thin input disk distribution (the latter depicted in black dashed line). At radii larger than $\sim0.2$~pc the apparent semi-major axis distribution significantly deviates from the initial input, where the $a_\inf$ is inferred to have much larger values than the input ones.  
As can be seen from Equation (\ref{eq:a}) large observed velocity yields a larger apparent semi-major axis of the stars around the SMBH than they actually are (left panel in Figure \ref{fig:SMAEcc}). 

 Note that the projected distance of the disk (i.e., $\sqrt{x^2+y^2}$) is independent on the binary velocity component.  In particular, as depicted in Figure \ref{fig:SMAEcc}, disk membership drops to $\sim 50\%$ for radii larger than the projected $6.5''$. The latter projected distance is often used in observation to determine the disk location. The results from our proof-of-concept calculations are consistent with the observed high disk membership inside $\sim 3.2''$ \citep[e.g.,][]{Yelda+14}. {In particular, we consider the $116$ stars with kinematic data from \citet{Yelda+14}, who estimated that about $20\%$ of the stars reside on the disk. We find the likelihood value, presented in \citet{Yelda+14}, that is associated with setting the sum over all fraction of stars in the disk, in each bin, to $20\%$, as 
 shown in the bottom panel of Figure \ref{fig:SMAEcc}, black line\footnote{{ We emphasize that we did not use \citet{Yelda+14} uncertainties estimations, instead we reduced their likelihood to assigned disk membership which allows to a ``cleaner'' comparison with our calculations.}  }. As depicted, the observations show the same sky projected trend as our simulations, where stars closer to the SMBH have larger {\it assigned} disk membership. This similarity in trend suggests that the binary fraction in the galactic center is rather large. In fact, taking this analysis at face value, the results presented here suggest that the disk fraction should be higher by a factor of $2.5$ (estimated roughly from the comparison between the red and black lines in Figure  \ref{fig:SMAEcc}). Thus, suggesting a possible {\it true} disk fraction of $\sim50\%$\footnote{The fraction of disk membership is estimated by summing over all of the bins, i.e., $\sum n \Delta a_2 / \Delta a_{\rm Disk}$, where $n$ is the number of stars in the disk inside of projected bin $ \Delta a_2$ (in linear scale). We then divid by the total linear projected length of the disk ($\Delta a_{\rm Disk}\sim 0.46$~pc). }.
  It is interesting to point out that the confirmed binary  IRS 16SW, which has a detected acceleration and which resides rather close to the inner edge of the SMBH (the first bin from the left, in the plot) is the only star at the inner disk edge with a low disk membership probability.}

The stars in the razor thin disk were set with $e_D=0$, but their apparent eccentricity distribution has a wide range with an average of $\sim0.23$ and standard deviation of $0.2$, taking only bound orbits. The probability distribution is depicted in Figure \ref{fig:SMAEcc}, right panel, solid blue line.  
Furthermore, we also found that about $3\%$ of the inferred systems resulted in eccentricities that are nearly radial or completely unbound on hyperbolic orbit. This may explain the observed high eccentricity and the nearly radial orbits reported by \citet{Bartko+09}. 
 Taking only the disk members in our simulation (i.e., either $\beta\leq 11.2^\circ$), we plot their eccentricity's probability distribution in Figure \ref{fig:SMAEcc}, top right panel. The average disk members' eccentricity is $0.14$ with standard deviation of $0.13$. 
Thus, our simulations suggest that perhaps some of the observed eccentric stars may actually have a circular orbit around the SMBH, and their eccentricity is only inferred to be high, as a result from the induced z-component velocity of a binary.  
 
 We also take \citet{Yelda+14} reported eccentricity at face value and repeat the above exercise for initial $e_D=0.3$. We found that all the above angles (i.e., $i_\inf$, $\Omega_\inf$, $\alpha$ and $\beta$) produce similar probability distribution, and thus are omitted from the Figures to avoid clutter. On the other hand the inferred eccentricity probability distribution is different (as depicted in Figure \ref{fig:SMAEcc}, green dashed lines), with an average of $0.4$ and a standard deviation of $0.24$, taking only bound orbits. About $3\%$ of systems  resulted on a hyperbolic orbits in this case.  As illustrate in this Figure, if the orbits around the SMBH were indeed eccentric, the binary contribution would have implied an apparent eccentricity, with an even higher values. This may explain the large eccentricities reported in the literature \citep[e.g.,][]{Bartko+09,Yelda+14}.

\begin{figure}
  \centering
 \hspace{-0.7cm}
 \includegraphics[width=1.1\linewidth]{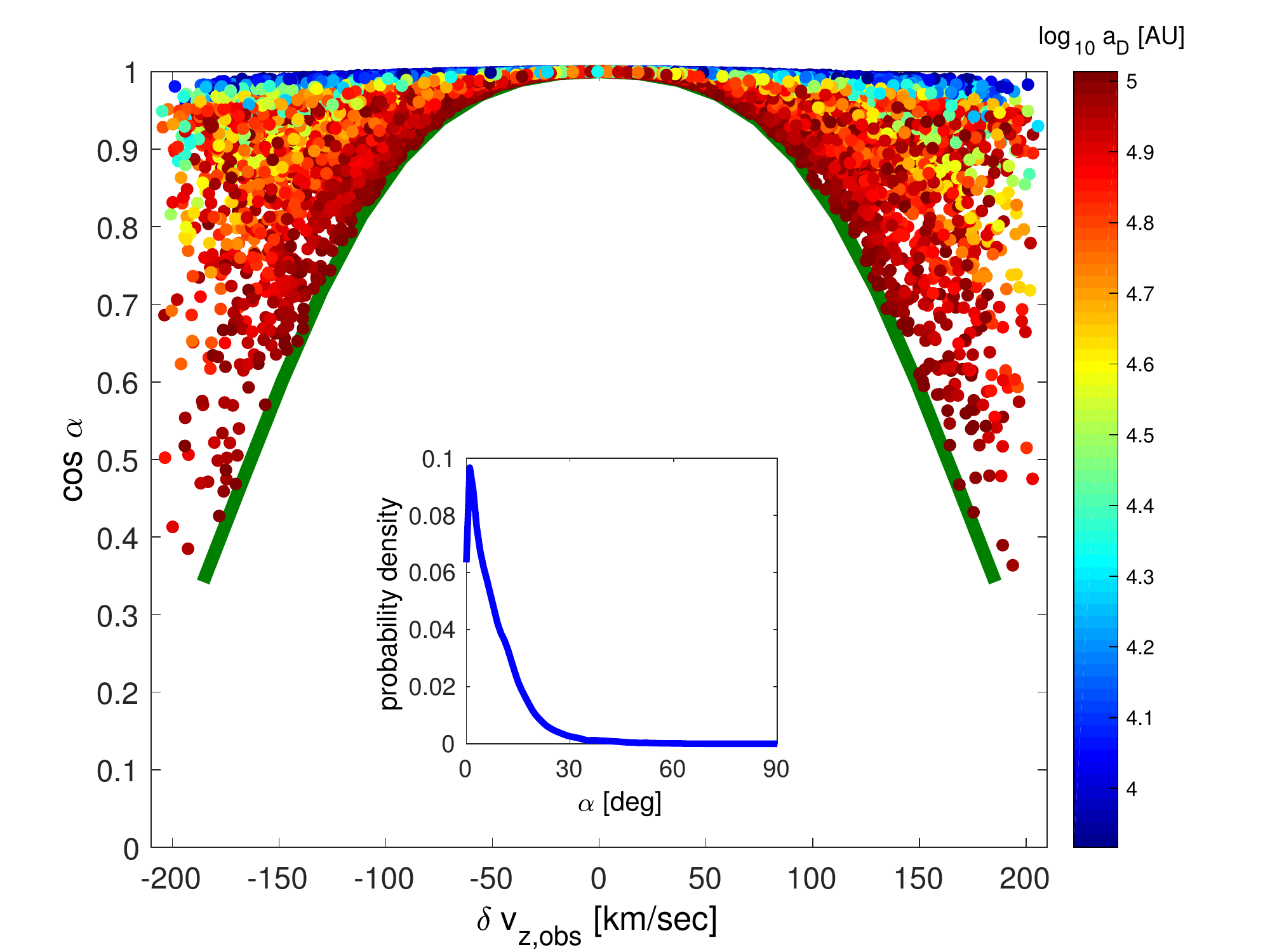}
    \caption{ The angle between the two angular momenta ($\alpha$) as a function of the binary radial velocity. The different color corresponds to the input semi-major axis of each binary around the SMBH in log space. The green solid line shows the analytical Equation (\ref{eq:cosalpha}). The inset show $\alpha$ distribution. }\label{fig:DiskvelSMA}
\end{figure}

While $\Omega$, $i$ and $\beta$ are sky projected quantities, we are also interested in the physical tilt between the input disk's angular momentum and the resulted apparent disk.  
Thus, we calculate the relative deflection angle $\alpha$ between the actual stellar disk's angular momentum ${\bf h}_D$ and the apparent  one ${\bf h}_\inf$ where 
\begin{equation}
\cos \alpha = \frac{ {\bf h}_D \cdot {\bf h}_\inf  }{h_D h_\inf}  \ , 
\end{equation}
The distribution of this angle 
is shown in the inset of Figure  \ref{fig:DiskvelSMA}.  This distribution peaks around $\sim 1^\circ$ with an average of $8^\circ$ with a standard deviation of $8^\circ$. In other words, a razor thin disk appears to be puffed when the mutual motion of the binaries is neglected. { In addition to that, the disk appears tilted compared to its actual orientation as the peak of $\alpha$ is not oriented at zero}.

The stars' Keplerian orbits imply that binaries that are further away from the disk, have smaller velocity around the SMBH compared to stars that are closer to the SMBH (see Figure \ref{fig:Diskvel}). Thus, the stars that are closer to the SMBH will have smaller deflection angle (i.e, $\cos \alpha \sim 1$) relative to the stars that are further away in the disk. We depict this behavior in Figure \ref{fig:DiskvelSMA}, were the larger deflection angles are associated with larger semi-major axes ($a_D$) from the SMBH. 
From vectorial identities, and using the fact that the binary velocity excess is observed only in the $z$-component (i.e., $\Delta=\delta  v_{z,\rm obs} /v_D$), we can write:
\begin{equation}\label{eq:cosalpha}
\cos\alpha= \frac{1+(v_z/v_D) \Delta}{\sqrt{ 1+2 (v_z/v_D)  \Delta +(x^2+y^2)/r_D^2   \Delta^2}} \ ,
\end{equation}
and in terms of the inclination and longitude of ascending nodes we can write:
\begin{equation}\label{eq:ab}
\cos\alpha= \sin i_D\sin i_\inf \cos(\Omega_D-\Omega_\inf) + \cos i_D\cos i_\inf \ .
\end{equation}
In Figure  \ref{fig:DiskvelSMA}, we show the analytical relation from Equation (\ref{eq:cosalpha}) for the far edge of the disk $0.5$~pc, and the maximum $v_z/v_D$ in the disk (which is  about $0.77$). Large $\alpha$ can be mistakenly interpreted as if stars lay outside the plane of the disk. As expected, the error in estimating the stellar disk's membership (small $\beta$ and thus small $\alpha$, from Eq.~(\ref{eq:ab})) is highly correlated with the distance from the SMBH. Stars that have wider orbit around the SMBH need larger $\delta v_{z,\obs}$ to produce larger variations in $\alpha$, and thus are less likely to be sensitive to binary configuration.

\section{Discussion and Conclusions}\label{sec:dis}

We have shown that the binarity nature of stars can significantly influence the apparent (observed) stellar disk properties in the GC. 
The stellar disk in the GC is estimated to be located between $0.04$~pc and $0.5$~pc, with sky projected inclination of $i_D=130^\circ$ ascending nodes of $\Omega_D=96^\circ$ \citep{Yelda+14}. Since the determination of disk membership is often estimated by one RV measurement, a velocity wobble on the z--axis (the line-of-sight) due to the binary motion can change the apparent properties.

We have conducted a simple proof-of-concept calculation that considered a razor-thin disk around the SMBH in the GC, assuming that each star is, in a binary configuration (see Figure \ref{fig:Diskvel}).   The apparent physical and projected orbital parameter are substantially different compared to the input one. For example, the inclination and ascending nodes deduced when ignoring the possibilities of an extra z-component that arises from the binary motion results in a  wide distribution for these angles (as depicted in Figure \ref{fig:angles}). The width of the distribution is largely caused by the velocity distribution, which is sensitive to the binary mass ratio and binary orientation (both drawn from a uniform distributions).  The binary motion causes the apparent location of the disk to be slightly shifted from the input location (as shown in Figure \ref{fig:angles} bottom panel). 
Note that we assumed a $100\%$ binary fraction in the disk, as we have fixed all values which are model dependent (e.g., binary fraction and binaries eccentricity and separation distribution). These effects may changed the quantitate results presented here (e.g., a distribution of binary separations may increase the effect) but not the qualitative effect.

 {As expected disk members that are farther away from the SMBH will be more sensitive to the binary motion than those that are closer to the SMBH since the latter have larger orbital velocity $v_D$.  The binary wobble velocity along the line of sight cause a system to appear off the disk and further away from the SMBH than truly is.  
In particular, we are more likely to deduce that stars that are closer to the SMBH belong to the disk, than those that are initially further away from the SMBH. This behavior is shown in Figure \ref{fig:SMAEcc}, bottom panel, and remarkably, this trend is consistent with observations, depicts as the black line in this Figure. Considering the observations, we find that the binary  IRS 16SW, which resides rather close to the SMBH, was found to have low probability to reside in the disk \citep{Yelda+14},  consistent with our finding.}

{We note that \citet{Yelda+14}  showed that observational sampling and prior assumptions can have large impact on the probability of disk membership. Using a forward modeling approach, they find that observations along the line of nodes of the disk along with uniform acceleration priors in the analysis can result in an over prediction for the disk by a factor of $2$. Future quantitative comparisons with this work will likely need to account for similar types of observational biases. }

Furthermore, the disk eccentricity may appear different than the actual orbital eccentricity. 
Starting with circular orbits around the SMBH, we found that the apparent eccentricity around the SMBH significantly differs from zero (see Figure \ref{fig:SMAEcc}, right panel). Interestingly, the apparent disk's eccentricity distribution (blue line in Figure \ref{fig:SMAEcc}) has a mean value of $\sim0.26$, which is consistent with \citet{Yelda+14} estimation from observations.  The apparent disk eccentricity can reach extreme values and even results in an apparent unbound orbits, which is also consistent with observations \citep[e.g.,][]{Bartko+09}.
Not only the disk's eccentricity will deviate from the physical one, but also the disk's members semi-major axis. Ignoring the contribution from binary motion will cause many disk members to appear more distant from the SMBH than they actually are (see Figure  \ref{fig:SMAEcc} left panel).

We defined a disk membership condition using a quantity that roughly describes an angular distance from the disk location on the sky, $\beta$, in the longitude of ascending nodes, $\Omega$ and the inclination, $i$, plane. We estimate the disk membership by considering the systems that lay within $\beta_{68}$ (i.e., corresponding to $68\%$ of systems around the initial location of the disk). As expected, we find that the disk fraction decreases as a function of the projected distance  (e.g., Figure \ref{fig:SMAEcc}, bottom panel). This trend is consistent with the observations, as shown in that Figure, black line. This functional agreement, suggests that binaries are more prevalent in the Galactic Center and may result in reducing the disk membership. Taking on face value our initial conditions and comparing our results to observations, may suggest that disk membership can be as high as $\sim 50\%$.

We conclude that the possibility of existing binaries in the measurements of the stellar disk properties cannot be ignored. As we showed, many of the puzzles and controversies involved the characteristics of the disk may be explained by the existence of binaries.
We encourage the community to take more measurements of the z-component of the stars' velocities. 

\acknowledgements
We thank the referee for his/her useful comments. 
S.N and A.M.G thank the Keck foundation for their partial support of the {\it NStarsOrbits} Project. 
S.N. acknowledges partial support from a Sloan Foundation Fellowship. A.M.G thanks the NSF grant number  AST-1412615.


\end{document}